\documentclass{iacrtrans}
\usepackage[utf8]{inputenc}
\usepackage{doi}
\usepackage{hyperref}

\author{Hong-Sheng Huang\inst{1} \and Yu-Lei Fu\inst{1,*} \and Han-Yu Lin\inst{2}}
\institute{Department of Computer Science and Engineering, National Taiwan Ocean University, Keelung, Taiwan, \email{00757205@email.ntou.edu.tw, 00657135@email.ntou.edu.tw, hanyu@email.ntou.edu.tw}}
\title[The Variant of Designated Verifier Signature Scheme with Message Recovery]{The Variant of Designated Verifier Signature Scheme with Message Recovery}
\subtitle{}

\begin{document}

\maketitle

\keywords[strong designated verifier signature \and message recovery \and universal designated verifier signature \and privacy-preserving]{strong designated verifier signature \and message recovery \and universal designated verifier signature \and privacy-preserving}

\begin{abstract}
  In this work, we introduce a strong Designated Verifier Signature (DVS) scheme that incorporates a message recovery mechanism inspired by the concept of the Universal Designated Verifier Signature (UDVS) scheme. It is worth noting that Saeednia's strong designated verifier signature scheme fails to guarantee the privacy of the signature, making it unsuitable for certain applications such as medical record certificates or voting systems. To overcome this limitation, we extend Lee's strong designated verifier signature with a message recovery scheme to develop a universal designated verifier signature scheme. This universal designated verifier scheme is crafted to safeguard the privacy of signature holders, ensuring that only designated verifiers can authenticate the true signer and recover the messages.
\end{abstract}

\section*{Introduction}

The digital signature finds application in various domains such as digital certificates, secure payment protocols, electronic voting, among others, providing essential attributes like integrity, authenticity, and non-repudiation. At its core, the digital signature serves to establish the identity of signers who utilize their private keys for signing. Any third party can then verify the signature using the corresponding public key of the signer. This ensures that only the possessor of the private key can generate a valid signature. Importantly, signers are unable to repudiate signatures generated by them subsequently.

In certain specialized scenarios such as electronic voting\cite{Ray01}\cite{Schoenmakers99} and Yao's Millionaires' problem\cite{Yao82}, it is imperative to safeguard the initiator's identity. Specifically, not everyone should be able to verify the signature, ensuring the signer's privacy. This concept traces back to the concept of undeniable signatures introduced by Chaum \& Antwerpen\cite{Chaum90}. In their scheme, signers have the option to select individuals allowed to verify signatures within the protocol. However, their approach necessitates both the confirmation and disavowal steps to rely on zero-knowledge properties, requiring interactive cooperation between parties. Consequently, the signer bears the burden of verification computation.

Henceforth, the concept of the designated verifier signature (DVS) scheme was introduced by Jakobsson et al. in 1996\cite{Jakobsson96}. Their scheme improves upon the undeniable signature by enabling the signer to convincingly provide evidence to a designated verifier in a non-interactive manner. This capability is achieved through the designated verifier's ability to produce a valid DVS using their private key, a property known as transcript simulation. A simulation signature is indistinguishable from a valid signature. Without this property, the designated verifier would be unable to persuade any third party of the authenticity of the DVS. Consequently, when a receiver receives a signature from the signer, they can be confident in its validity. This property makes it challenging for any third party to ascertain the true signer and efficiently verify the authenticity of the DVS.

However, Saeednia et al.\cite{Saeednia03} identified a potential issue where a third party could be highly likely to believe that a signature originates from the signer, especially if they obtain the signature before the designated verifier receives it. To address this concern, they proposed an efficient designated verifier signature scheme. This scheme achieves the property of strongness without relying on public verifiability by incorporating the designated verifier's private key into the verification step. As a result, any third party lacking the designated verifier's private key would be unable to verify the signature. Furthermore, they would not be able to access any message contained within the signature.

Some signature schemes, such as those discussed by Nyberg and Ruepple\cite{Nyberg93}\cite{Nyberg94}, incorporate a message recovery feature. Rather than embedding the message within the signature, they integrate the message as part of the signature itself. This allows designated verifiers to simultaneously recover the message and verify the signature. Additionally, Lee and Chang\cite{Lee07} propose a robust designated verifier signature scheme that enhances the message recovery capability. Their scheme ensures that third parties cannot identify the real signer nor access the original message.

Since the proposal of the designated verifier signature (DVS), numerous researchers have delved into its study and developed various variants. One notable variant is the universal designated verifier signature (UDVS) scheme, initially introduced by Steinfeld et al.\cite{Steinfeld03}\cite{Steinfeld04}. In this scheme, the signer and the signature holder are distinct individuals, enhancing the privacy of the signer's identity. It enables any signature holder to designate a verifier's public key non-interactively, facilitating publicly verifiable signatures. The designated verifier can then validate the UDVS using their private key but is unable to transfer the evidence to any third party, thereby achieving non-transferability.

Furthermore, several variants of the UDVS rely on different assumptions. For instance, Zhang et al.\cite{Zhang05} proposed a UDVS scheme based on the strong Diffie-Hellman problem (SDHP), while Huang et al.\cite{Huang08} introduced a UDVS scheme based on the gap bilinear Diffie-Hellman problem (GBDHP). Moreover, Han-Yu Lin\cite{Lin13} proposed another UDVS based on the bilinear inverse Diffie-Hellman problem (BIDHP) and extended it to the universal designated multi-verifier signature (UDMVS).

Our work extends the strong Designated Verifier Signature scheme proposed by Lee and Chang, drawing inspiration from the concept of universal designated verifier signature collaboration for message recovery. This extension ensures privacy for both the signer and designated verifiers. Such an approach holds significant promise in applications such as certificate management for medical records and electronic voting systems.

\section{Preliminaries}
In this section, we briefly describe the security notations and the schemes used.

\subsection{Strong Designated Verifier Signature Scheme}
\subsubsection{Setup}
Let $p$ and $q$ be two large primes such that $q | p-1$ and $g$ be an element of $\mathbb{Z}_{p}^{*}$ of order $q$. The message $m \in \mathbb{Z}_{p}$. Alice's public key is $y_{A} = g^{x_{A}} \> mod \> p$, where $x_{A} \in \mathbb{Z}_{q}^{*}$ is her secret key, and Bob's public key be $y_{B} = g^{x_{B}} \> mod \> p$, where $x_{B} \in \mathbb{Z}_{q}^{*}$ is his secret key. One-way hash function $H$ outputs values in $\mathbb{Z}_{q}$.
\subsubsection{Signature Generation} Alice wants to send strong designated verifier signature $(r, s, t)$ with a message $m$ to Bob. Alice choose two random number $k \in \mathbb{Z}_{q}$ and $t \in \mathbb{Z}_{q}^{*}$, and then generate the signature,
\begin{align*}
    c &= y_{B}^{k} \> mod \> p\\
    r &= H(m, c)\\
    s &= (kt^{-1}-rx_{A}) \> mod \> q
\end{align*}
\subsubsection{Signature Verification} After receiving the transcript $(r, s, t)$ with message $m$, Bob would verifies the signature
\begin{align*}
    H(m, (g^{s}y_{A}^{r})^{tx_{B}} \> mod \> p) \stackrel{\text{\tiny ?}}{=} r
\end{align*}
Obviously, nobody can perform this verification except Bob, since Bob's secret key is involved in the verification equation.
\subsubsection{Correctness}
\begin{align*}
    &H(m, (g^{s}y_{A}^{r})^{tx_{B}} \> mod \> p) \\ &= H(m, (g^{x_{B}}y_{A}^{r})^{st} \> mod \> p) \\ &= H(m, (g^{x_{B}}g^{x_{A}r})^{st} \> mod \> p) \\ &= H(m, (g^{x_{B}k} \> mod \> p)) \\ &= H(m, (y_{B}^{k} \> mod \> p)) \\ &= H(m, c) \\ &= r
\end{align*}
\subsubsection{Transcript Simulation}
Bob select $s' \in \mathbb{Z}_{q}$ and $r' \in \mathbb{Z}_{q}^{*}$ at random and computes the simulated signature
\begin{align*}
    c &= g^{s'}y_{A}^{r'} \> mod \> p \\
    r &= H(m, c)\\
    \ell &= r'r^{-1} \> mod \> q\\
    s &= s'\ell^{-1} \> mod \> q\\
    t &= \ell x_{B}^{-1} \> mod \> q
\end{align*}
Bob can simulate a signature with $m$ as $(r, s, t)$. If Bob's secret key is shared with a third party, that party can verify the signature similarly to Bob. However, because Bob can generate the transcript in an indistinguishable manner as described above, the third party cannot efficiently discern the true signer of that signature.

\subsection{Strong Designated Verifier Signature Scheme with message recovery mechanism}
\subsubsection{Setup}
Let $p$ and $q$ be two large primes such that $q | p-1$ and $g$ be an element of $\mathbb{Z}_{p}^{*}$ of order $q$. The message $m \in \mathbb{Z}_{p}$. Alice's public key is $y_{A} = g^{x_{A}} \> mod \> p$, where $x_{A} \in \mathbb{Z}_{q}^{*}$ is her secret key, and Bob's public key be $y_{B} = g^{x_{B}} \> mod \> p$, where $x_{B} \in \mathbb{Z}_{q}^{*}$ is his secret key. One-way hash function $H$ outputs values in $\mathbb{Z}_{q}$.
\subsubsection{Signature Generation}
Alice wants to generate a strong designated verifier signature $(t, c, r, s)$ with message recovery to Bob. Alice choose two random numbers $k_{1}$ from $\mathbb{Z}_{q}^{*}$ and $k_{2}$ from $\mathbb{Z}_{q}$ then generate a signature $(t,c,r,s)$
\begin{align*}
    t &= g^{k_{1}} \> mod \> p\\
    c &= my_{B}^{k_{2}} \> mod \> p\\
    r &= H(m, g^{k_{2}})\\
    s &= (k_{1}^{-1}(x_{A}r-k_{2})) \> mod \> q
\end{align*}
\subsubsection{Message Recovery and Verification}
After Bob receiving signature $(t, c, r, s)$ from Alice, Bob simultaneously recover the message and verifies the signature
\begin{align*}
    m &= c(t^{s}y_{A}^{-r})^{x_{B}} \> mod \> p\\
    r &\stackrel{\text{\tiny ?}}{=} H(m, y_{A}^{r}t^{-s})
\end{align*}
\subsubsection{Correctness}
\begin{align*}
    m &= c(t^{s}y_{A}^{-r})^{x_{B}} \> mod \> p\\
    &= c(g^{x_{B}sk_{1}}g^{x_{A}(-r)}) \> mod \> p\\
    &= c(g^{x_{B}sk_{1}-x_{A}r}) \> mod \> p\\
    &= c(g^{x_{B}-k_{2}}) \> mod \> p \\
    &= cy_{B}^{-k_{2}} \> mod \> p\\
    &= my_{B}^{k_{2}} \> mod \> p\\
    &= c\\
    r &= H(m, y_{A}^{r}t^{-s})\\
    &= H(m, g^{x_{A}r}g^{k_{1}(-s)})\\
    &= H(m, g^{x_{A}r-sk_{1}})\\
    &= H(m, g^{k_{2}}) \\
    &= r
\end{align*}
\subsubsection{Transcript Simulation}
Bob can simulate the designated verifier signature $(t, c, r, s)$ with message $m$, Bob selects two random values $w_{1} \in \mathbb{Z}_{q}^{*}$ and $w_{2} \in \mathbb{Z}_{q}$. Then he simulate $(t, c, r, s)$
\begin{align*}
    t &= y_{A}^{w_{1}^{-1}} \> mod \> p\\
    c &= (my_{A}^{x_{B}w_{1}^{-1}w_{2}}) \> mod \> p\\
    r &= H(m, y_{A}^{w_{1}^{-1}w_{2}})\\
    s &= (w_{1}r-w_{2}) \> mod \> q
\end{align*}

\section{Proposed Scheme}
We outline the involved parties and the composed algorithms of our proposed strong designated verifier signature scheme, inspired by the universal designated verifier signature scheme, and subsequently provide a detailed construction.
\subsection{Involved Parties}
A universal designated verifier signature scheme has two involved parties: a signer and a verifier. Each one is a probabilistic polynomial-time Turing machine (PPTM). The signer generates a publicly verifiable signature (PV-signature) such that the verifier can validate it with signer’s public key.
\subsection{Algorithms}
The proposed scheme consists of three algorithms (including Setup, PSG and PSV). We describe these algorithms as follows:
\subsubsection{Setup}
Taking as input $k_{1}$ from $Z_{q}^{*}$ and $k_{2}$ from $Z_{q}$ where $k_{1}, k_{2}$ are security parameters, the algorithm generates the system’s public parameters $params$.
\subsubsection{PV-Signature-Generation (PSG)}
The PSG algorithm takes as input the system parameters params, a message and the private key of signer. It generates a PV-signature $\Omega$.
\subsubsection{PV-Signature-Verification (PSV)}
The PSV algorithm takes as input the system parameters $params$, a PV-signature $\Omega$ along with the corresponding message $m$, and the public key of signer. It outputs True if $\Omega$ is a valid PV-signature for $m$. Otherwise, the error symbol $\P$ is returned as a result.
\subsection{Construction of Probabilistic Signature Scheme}
We detail the construction of our probabilistic signature scheme
\subsubsection{Setup}
Let $p$ and $q$ be two large primes such that $q | p-1$ and $g$ be an element of $Z_{p}^{*}$ of order $q$. The message $m \in Z_{p}$. Alice's public key is $y_{A} = g^{x_{A}} \> mod \> p$, where $x_{A} \in Z_{q}^{*}$ is her secret key, and Bob's public key be $y_{B} = g^{x_{B}} \> mod \> p$, where $x_{B} \in Z_{q}^{*}$ is his secret key. One-way hash function $H$ outputs values in $Z_{q}$.
\subsubsection{PV-Signature-Generation (PSG)}
Let signer choose two random integer $k_{1} \in Z_{q}^{*}$ and $k_{2} \in Z_{q}$ generates signature $(t, c, r, s)$
\begin{align*}
    t &= g^{k_{1}} \> mod \> p\\
    c &= mg^{k_{2}} \> mod \> p\\
    r &= H(m, g^{k_{2}})\\
    s &= k_{1}^{-1}(x_{A}r - k_{2}) \> mod \> q
\end{align*}
The PV-signature $\Omega = (t, c, r, s)$ with message $m$.
\subsubsection{PV-Signature-Verification (PSV)}
To check the validity of the PV-signature $\Omega = (t, c, r, s)$, anyone can verify whether
\begin{align*}
    m &\stackrel{\text{\tiny ?}}{=} c(t^{s}y_{A}^{-r}) \> mod \> p\\
    r &\stackrel{\text{\tiny ?}}{=} H(m, t^{-s}y_{A}^{r})
\end{align*}

\subsubsection{Correctness}
\begin{align*}
    m &= c(t^{s}y_{A}^{-r} \> mod \> p) \\
    &= c(g^{sk_{1}}g^{-x_{A}r}) \> mod \> p\\
    &= c(g^{sk_{1}-x_{A}r} \> mod \> p\\
    &= c(g^{-(-sk_{1}+x_{A}r)}) \> mod \> p\\
    &= cg^{-k_{2}} \> mod \> p\\
    &= mg^{k_{2}} \> mod \> p\\
    &= c\\
    r &= H(m, t^{-s}y_{A}^{r})\\
    &= H(m, g^{-sk_{1}}g^{x_{A}r})\\
    &= H(m, g^{-sk_{1}+x_{A}r})\\
    &= H(m, g^{k_{2}})\\
    &= r
\end{align*}

\section{Extension into UDVS Scheme}
In this section, we present DVS scheme based on the UDVS scheme. We first address involved parties and composed algorithms of our UDVS scheme and then give concrete construction.
\subsection{Involved parties}
A conventional UDVS scheme has three involved parties including a signer, a designator (signature holder) and a designated verifier. In our proposed UDVS schemes, each party is a probabilistic polynomial-time Turing machine (PPTM). The signer will generate a PV-signature and send it along with the message to the designator. After validating the PV-signature, the designator further creates a designated verifier signature (DV-signature) and deliveries it together with the message to the designated verifier. Consequently, the DV-signature can only be verified by the designated verifier with his private key. Besides, the designated verifier can not transfer the conviction to any third party, since he is also capable of generating another computationally indistinguishable transcript.
\subsection{Algorithms}
The proposed UDVS scheme consists of five algorithms (including Setup, PSG, PSV, DSG and DSV). The first three algorithms are defined the same as those in our probabilistic signature scheme
\subsubsection{DV-Signature-Generation (DSG)}
The DSG algorithm takes as input a PV-signature $\Omega$ along with the corresponding message $m$, and the public key of designated verifier. It generates a DV-signature $\delta$.
\subsubsection{DV-Signature-Verification (DSV)}
The DSV algorithm takes as input a DV-signature $\delta$ along with the corresponding message $m$, the private key of the designated verifier, and the public key of signer. It outputs $\textbf{True}$ if $\delta$ is a valid DV-signature for $m$. Otherwise, the error symbol $\P$ is returned as a result.
\subsection{Concrete construction of UDVS scheme}
We demonstrate the proposed UDVS scheme in the subsection. This scheme is a conventional UDVS which only allows the signature holder to solely designate the PV-signature to one intended designated verifier without further interactions.
\subsubsection{DV-Signature-Generation (DSG)}
After receiving $\Omega = (t, c, r, s)$ of $m$, let designator choose a random number $d \in Z_{q}$ and compute
\begin{align*}
    e &= g^{-d} \> mod \> p\\
    w &= cy_{B}^{d}
\end{align*}
Then deliver the $\delta = (t, w, r, s, e)$ to the designated verifier along with the corresponding message $m$.
\subsubsection{DV-Signature-Verification (DSV) with message recovery}
Upon receiving $(\delta , m)$, the designated verifier verifies whether
\begin{align*}
    m &\stackrel{\text{\tiny ?}}{=} w(t^{s}y_{s}^{-r}e^{x_{v}}) \> mod \> p\\
    r &\stackrel{\text{\tiny ?}}{=} H(m, t^{-s}y_{s}^{r})
\end{align*}
\subsubsection{Correctness of DV-Signature-Verification (DSV) with message recovery}
\begin{align*}
    m &= w(t^{s}y_{A}^{-r}e^{x_{B}}) \> mod \> p \\
    &= w(g^{sk_{1}-x_{A}r+x_{B}(-d)})\\
    &= w(g^{-(-sk_{1}+x_{A}r)}g^{x_{B}(-d)})\\
    &= wg^{-k_{2}}y_{B}^{-d} \\
    &= mg^{k_{2}}y_{B}^{d} \\
    &= cy_{B}^{d}\\
    &= w\\
    r &= H(m, t^{-s}y_{A}^{r}) \\
    &= H(m, g^{-sk_{1}}g^{x_{A}r})\\
    &= H(m, g^{-sk_{1}+x_{A}r})\\
    &= H(m, g^{k_{2}})\\
    &= r
\end{align*}
\subsubsection{Transcript simulation}
Let designated verifier selects $w_{1} \in Z_{q}^{*}$, $w_{2} \in Z_{q}$ and $d' \in Z_{q}$ at random and compute $(t', w', r', s', e')$ the simulated signature
\begin{align*}
    t' &= y_{A}^{w_{1}^{-1}} \> mod \> p\\
    c' &= (my_{A}^{x_{B}w_{1}^{-1}w_{2}}) \> mod \> p\\
    r' &= H(m, y_{A}^{w_{1}^{-1}w_{2}}) \> mod \> p\\
    s' &= (w_{1}r - w_{2}) \> mod \> q\\
    e' &= cy_{B}^{d'}
\end{align*}

\section{Analysis}
In this section, we analyze our proposed scheme, which is a designated verifier signature, inspired by the concept of the universal designated verifier signature. Additionally, we establish that it is unforgeable for any third party without access to either the signer's secret key or the recipient's secret key.

\subsection{Strong designated verifier property}
Our scheme is a designated verifier signature scheme. To demonstrate this, we establish that the transcripts simulated by the designated verifier are indistinguishable from those generated by the signer. To simulate a signature, the designated verifier randomly selects $w_{1}$ from $Z_{q}^{*}$, $w_{2}$ from $Z_{q}$, and $d'$ from $Z_{q}$. The simulated signature $(t', c', r', s', e')$ is then generated from these three values. The probability that this simulated transcript by the designated verifier represents a signature randomly chosen from the set of all possible signer's signatures is $\frac{1}{q(q-1)(q-2)}$. Therefore, both signatures follow the same probability distribution, confirming that the proposed scheme is indeed a designated verifier signature scheme.

Moreover, the proposed scheme satisfies the strongness property by involving the designated verifier's secret key in both the message recovery and verification steps. Thus, upon observing a signature $(t, c, r, s)$, no one can discern the real signer or recover the message except the designated verifier. If the designated verifier discloses their secret key to the public, anyone can then recover the message and verify the signature. However, even in this scenario, no one can conclusively determine whether the signature originates from the signer or the designated verifier.

\subsection{Unforgeability}
While the signature should be forgeable by the designated verifier, it must not be forgeable by any third party. The signer does not generate the value of $t$ with his public key. Two scenarios can be considered for potential forging attempts by an attacker.

Firstly, the attacker may attempt to generate a signature as the signer does. They would randomly choose $k_{1}$ from $Z_{q}^{*}$, $k_{2}$ from $Z_{q}$, and compute $t$, $c$, $r$ and $s$ using the relevant formulas. Next, the attacker would aim to find $s$ that satisfies the verification step. This requires knowledge of the signer's secret key.

Secondly, the attacker may try to simulate a signature as the designated verifier does. They would randomly select $w_{1}$ from $Z_{q}^{*}$, $w_{2}$ from $Z_{q}$, $d$ from $Z_{q}$, and attempt to compute $t'$, $c'$, $r'$, $s'$, and $e'$. Since the designated verifier's secret key is necessary to compute $c'$, it becomes infeasible for the attacker to compute $c'$. In both scenarios, a successful forgery by any third party implies that the attacker has solved the discrete logarithm problem.

\subsection{Confidentiality}
Confidentiality means that only the designated verifier can recover the message. It is evident that the proposed scheme satisfies confidentiality since the designated verifier's secret key is required to recover the message.

\section{Conclusion}
In this work, we propose variants of a strong designated verifier signature scheme with a message recovery mechanism, inspired by the concept of the universal designated verifier signature scheme. While the security assumption of our scheme still relies on the discrete logarithm problem, we incorporate the concept of UDVS, which separates the roles of the signer, designator, and designated verifier. This abstraction enhances the obfuscation of our mechanism, allowing us to achieve unforgeability and confidentiality.


\end{document}